\documentclass[prd,preprintnumbers,nofootinbib]{revtex4} 
\usepackage{graphicx} 
\usepackage{amsmath}
\usepackage{amsfonts,amsbsy}
\usepackage{amssymb}

\def\be{\begin{equation}}
\def\ee{\end{equation}}
\def\bea{\begin{eqnarray}}
\def\eea{\end{eqnarray}}

\def\eq#1{{Eq.~(\ref{#1})}}
\def\fig#1{{Fig.~\ref{#1}}}

\begin{document}

\title{\bf Hadron production in pA collisions at the LHC from the Color Glass Condensate 
}


\author{Jamal Jalilian-Marian$^{1,2}$ and Amir H. Rezaeian$^{3}$}
\affiliation{
$^1$ Department of Natural Sciences, Baruch College, CUNY,
17 Lexington Avenue, New York, NY 10010, USA\\
$^2$ The Graduate School and University Center, City
  University of New York, 365 Fifth Avenue, New York, NY 10016, USA\\
$^3$Departamento de F\'\i sica, Universidad T\'ecnica
Federico Santa Mar\'\i a, Avda. Espa\~na 1680,
Casilla 110-V, Valparaiso, Chile }

\begin{abstract}
We investigate the contribution of inelastic and elastic processes to
single inclusive hadron production in proton-proton and proton
(deuteron)-nucleus collisions at RHIC and the LHC. Using the hybrid
formulation which includes both elastic and inelastic contributions,
supplemented with the running-coupling Balitsky-Kovchegov equation,
we get a good description of RHIC data. It is shown
that inclusion of the inelastic terms makes the transverse momentum
dependence of the production cross section steeper in the mid-rapidity
region but does not affect the cross section in the very forward
region. The inelastic processes also lead to a sharper increase of the
nuclear modification factor $R_{pA}$ with increasing $p_T$. We also
make predictions for the nuclear modification factor in proton-nucleus
collisions at the LHC ($\sqrt{s}=4.4$ and $8.8$ TeV) at various
rapidities using the Color Glass Condensate framework.
\end{abstract}

\maketitle

\section{Introduction}
The Color Glass Condensate (CGC) formalism~\cite{review} is a self-consistent, effective theory approach 
to QCD interactions at high energy (or equivalently small $x$). Even though it is a weak coupling
approach, it is different from the collinear factorization based approach of pQCD
in two important aspects; first, it re-sums quantum corrections which are enhanced
by large logarithms of $1/x$ as opposed to large logarithms of $Q^2$ in pQCD and second, 
it includes high gluon density effects which are important at small $x$ and/or for large
nuclei where the physics of gluon saturation may be the dominant.

The CGC formalism has successfully been applied to many QCD processes,
from fully inclusive ones such as structure functions in DIS to single
and double inclusive particle production in proton-proton and
proton-nucleus collisions at high energy, see Ref.~\cite{review} and references therein. The CGC formalism has been
also quite successful in providing predictions for particle multiplicities 
at the LHC \cite{a00,jav0,a000} and  may provide a first-principle way of understanding
of isotropization and thermalization of QCD matter produced in high
energy heavy ion collisions at RHIC and the LHC \cite{the}.

The observed suppression of single inclusive hadron production and the
disappearance of the away side peak in double hadron production in the
forward rapidity region of deuteron-nucleus (dA) collisions at
RHIC~\cite{exp,exp1} are perhaps the strongest evidence for the importance
and possibly dominance of saturation effects at RHIC. This will soon
be further tested at the LHC where one will be able to probe CGC
dynamics in a much larger kinematic region due to the larger energy of
the collisions at the LHC. Single inclusive hadron production in proton-nucleus (pA)
collisions at RHIC and the LHC has been investigated by many
authors~\cite{single,dhj,dhj2,mac,ja-cm,AS} in the CGC formalism using
varying degrees of approximations and models (for an alternative
description, see for example Refs.~\cite{pQCD,all}). The
most important ingredient of the single inclusive hadron production
cross section which captures the saturation dynamics is the
fundamental (or adjoint) dipole cross section, the imaginary part of
the quark anti-quark scattering amplitude on a proton or nucleus
target. This dipole cross section satisfies the JIMWLK/BK evolution
equations~\cite{jimwlk,bk} and re-sums the small $x$ as well as high
gluon density effects. The evolution equation for the dipole cross
section is now known with next-to-leading-order (NLO) accuracy~\cite{nlo}, see also Ref.~\cite{nlo-c}.  

There are two distinct but related approaches to hadron production in
high energy asymmetric (such as proton-nucleus or very forward
proton-proton and nucleus-nucleus) collisions. One is the well-known
$k_T$ factorized approach \cite{KTINC,kt-rest} where partons in both 
the projectile and
target are assumed to be at very small $x$ ($ x< 0.01$) so that the CGC
formalism is applicable to both the projectile and target. This
approach is valid as long as one stays away from the projectile
fragmentation region. An alternative approach was developed
in~\cite{dhj} where one treats the projectile wave-function
perturbatively, i. e. using the standard DGLAP picture while treating
the target by CGC methods. This approach is better suited for the
projectile fragmentation region. Very recently this approach has been improved by keeping the
inelastic pieces of the cross section which may be important at high
transverse momentum~\cite{ak}. Here we numerically investigate the
contribution of theses inelastic contributions to single inclusive
hadron production in proton-nucleus and proton-proton collisions at
RHIC and the LHC and show that inclusion of these terms improves the high
$p_T$ behavior of the cross section. The nuclear modification factor
is also shown to increase faster with increasing $p_T$ than the case
where these inelastic contributions are ignored.  

The paper is organized as follows: In the next section, we introduce
our formalism for the inclusive hadron production in pA
collisions, namely the hybrid formulation which includes both elastic and inelastic contributions,  
supplemented with the running-coupling Balitsky-Kovchegov equation. Section III is
devoted to comparison with the experimental data and to a discussion of
various predictions for the LHC energies.  We conclude in
Sec. IV.

\section{Single inclusive hadron production; main formulation} 

The cross section for single inclusive hadron production in asymmetric
collisions (scattering of a dilute system of parton with a dense one) at high energy is given by~\cite{ak}
\begin{eqnarray}\label{final}
\frac{dN^{p A \rightarrow h X}}{d^2p_T d\eta}&=&\frac{K}{(2\pi)^2}\Bigg[\int_{x_F}^1 \frac{dz}{z^2} \Big[x_1f_g(x_1,Q^2)N_A(x_2,\frac{p_T}{z})D_{h/g}(z,Q)+\Sigma_qx_1f_q(x_1,Q^2)N_F(x_2,\frac{p_T}{z})D_{h/q}(z,Q)\Big]\nonumber\\
&+&\int_{x_F}^1 \frac{dz}{z^2}\frac{\alpha_s}{ 2\pi^2}\frac{z^4}{p_T^4}\int_{k_T^2<Q^2}d^2k_T k_T^2 N_F(k_T,x_2)\int_{x_1}^1\frac{d\xi}{\xi}\Sigma_{i,j=q,\bar q, g}w_{i/j}(\xi)P_{i/j}(\xi)x_1f_j(\frac{x_1}{\xi}, Q)D_{h/i}(z,Q)\Bigg],
\end{eqnarray}
where $f_j(x,Q^2)$ is the parton distribution functions (PDF) of the
incoming proton which depends on the light-cone momentum fractions $x$
and the hard scale $Q$. The function $D_{h/i}(z,Q)$ is the
hadron fragmentation function (FF) of $i$`th parton to the final
hadron $h$ with a momentum fraction $z$.  The variables $\eta$ and $p_T$ are the
pseudo-rapidity and transverse momentum of the produced hadron. The
longitudinal momentum fractions $x_1$ and $x_2$ are defined as follows
\begin{equation}\label{xs}
x_F\approx \frac{p_T}{\sqrt{s}}e^\eta; \ \ \ \ x_1=\frac{x_F}{ z}; \ \ \ \ \ x_2=x_1e^{-2\eta},
\end{equation}
where $\sqrt{s}$ is the collision energy per nucleon. Here we
neglect hadron masses since we are only interested in light
hadron production at high-$p_T$ (thereby rapidity and 
pseudo-rapidity are equal). 

It is perhaps useful to remind the reader of the derivation of
Eq.~(\ref{final}).  The first line of Eq.~(\ref{final}) was first
derived in Ref.~\cite{dhj}. The result of \cite{dhj} has been recently
improved in \cite{ak} by keeping the inelastic pieces which lead to
the second line in Eq.~(\ref{final}).  Our main goal in this paper is
to consider the effect of this new term in the inclusive hadron
production at both RHIC and the LHC. Let us first focus on the first
line of Eq.~(\ref{final}), the DHJ term \cite{dhj}. We
refer the reader to Ref.~\cite{dhj} for technical details and just outline
the derivation of the DHJ term. One starts by calculating two particle
production cross section in proton-nucleus scattering. The simplest
process is when a quark from the projectile scatters on the target and
radiates a gluon either before or after the
scattering (see Fig. (12) in~\cite{jjm-yk}). The incoming quark as well as the outgoing
quark and the radiated gluon can all multiply scatter on the
target. However, if one is interested in single inclusive production,
one needs to integrate over one of the final state partons. Some of
the Feynman diagrams will have collinear divergences whereas others do
not.  There is a collinear divergence in the final state which happens
when the outgoing quark and the radiated gluon are nearly parallel and
only the initial state quark multiply scatters on the target. This
divergent term is absorbed into quark-hadron fragmentation function
and lead to its evolution with $Q^2$ according to the LO DGLAP evolution
equation. The finite (non-collinear divergent) pieces are ignored as
they are part of the NLO corrections.

There is also a collinear divergence in the initial state which happens when the incoming 
quark and the radiated gluon are nearly parallel and only the final state quark multiply 
scatters on the target. This sort of collinear divergence is absorbed into the incoming
parton distribution function and leads to its evolution according to the standard DGLAP 
evolution equation. Again, the finite parts of these terms were ignored in~\cite{dhj}
since they correspond to higher order (in $\alpha_s$) corrections. It was pointed out 
in \cite{ak} that the finite pieces which are ignored in \cite{dhj} may be important at high $p_T$
and therefore can lead to a modification of the production cross section. Keeping  
the finite diagrams (which do not have a collinear divergence) and making the high 
$p_T$ approximation (gradient expansion of the quadrupole cross section) leads to
Eq. (\ref{final}). While the above argument is for an incoming quark scattering on the
target, inclusion of other processes such as an incoming gluon scattering on the target
is straightforward~\cite{dmxy} where a similar analysis of the collinear divergences 
can be made. 

Now that the origin of these inelastic terms is made more clear, we comment on the relative
significance of the two contributions. The first piece of eq. (\ref{final}), dubbed the 
elastic part, corresponds to an incoming parton in the proton wave function scattering
elastically on the target. This incoming parton initially has zero transverse momentum but 
picks up transverse momentum of order $Q_s$ after multiply scattering on the target. This 
term should be most important when the transverse momentum of the produced hadron is of order
$Q_s$ or perhaps even a bit larger. The second term in eq. (\ref{final}), dubbed the inelastic
piece, corresponds to a high transverse momentum parton radiated from the incoming parton 
in the projectile wave function. This radiated parton is already at high transverse momentum and interacts
with the target only once (higher number of scatterings will be power suppressed). This term
is therefore important only when the produced hadron is at transverse momenta much higher that
the saturation scale of the target $Q_s$.

In Eq.~(\ref{final}), we have 
introduced a $K$-factor to mimic the effect of higher order corrections.  The inelastic weight functions 
$w_i/j$ are given by
\begin{eqnarray}\label{weights}
w_{g/g}(\xi)&=&2\frac{N_c^2}{ N_c^2-1}(1-\xi+\xi^2),\\
w_{g/q}(\xi)&=&w_{g/\bar q}(\xi)=\frac{N_c^2}{ N_c^2-1}\bigg[1+(1-\xi)^2-\frac{\xi^2}{N_c^2}\bigg],\\
w_{q/q}(\xi)&=&w_{\bar q/\bar q}(\xi)=\frac{N_c^2}{ N_c^2-1}\bigg[1+\xi^2-\frac{(1-\xi)^2}{N_c^2}\bigg],\\
w_{q/g}(\xi)&=&w_{\bar q/g}(\xi)=\frac{1}{2}\bigg[(1-\xi)^2+\xi^2-\frac{2\xi(1-\xi)}{N_c^2-1}\bigg],
\end{eqnarray}
where $N_c$ denotes the number of colors. The function $P_{i/j}$ in
\eq{final} denotes the Altarelli-Parisi splitting function that
describes the probability of a given parton $j$ splitting into two
others. The leading-order splitting functions (for $N_c=3$)
are given by \cite{pij}, 
\begin{eqnarray}\label{sp}
P_{q/q}(\xi)&=&\frac{4}{3}\bigg[\frac{1+\xi^2}{(1-\xi)_{+}}\bigg]+2\delta(1-\xi),\\
P_{qg}(\xi)&=&\frac{1}{2}\bigg[\xi^2+(1-\xi)^2\bigg],\\
P_{gq}(\xi)&=&\frac{4}{3}\bigg[\frac{1+(1-\xi)^2}{\xi}\bigg],\\
P_{gg}(\xi)&=& 6\bigg[\frac{1-\xi}{\xi}+\xi(1-\xi)+\frac{\xi}{(1-\xi)_{+}}\bigg]
+\left(\frac{11}{2}-\frac{n_f}{3}\right)\delta(1-\xi),\
\end{eqnarray}
where $n_f$ is the number of active flavor and subscript $+$ refers to
the so-called $''+''$ prescription used to regularize the singularities
as $\xi \rightarrow 1$ \cite{pij}. In
\eq{final}, the amplitude $N_F$ ($N_A$) is the two-dimensional Fourier
transform of the imaginary part of the forward dipole-target
scattering amplitude $\mathcal{N}_{A(F)}$ in the fundamental (F) or adjoint (A)
representation,
\begin{equation}
N_{A(F)}(x,k_T)=\int d^2\vec r e^{-i\vec k_T.\vec r}\left(1-\mathcal{N}_{A(F)}(r,Y=\ln(x0/x))\right),
\end{equation}
where $r=|\vec r|$ is the dipole transverse size. In the large-$N_c$
limit, one has the following relation between the adjoint and fundamental dipoles, 
\begin{equation}
\mathcal{N}_A(r,Y)=2\mathcal{N}_F(r,Y)-\mathcal{N}_F^2(r,Y). 
\end{equation}
 The amplitude $\mathcal{N}_{A(F)}$ incorporates all multi-scatterings
 between a projectile color-dipole and the target and encodes the small-x
 dynamics. In the CGC framework, it can be obtained from the solution of JIMWLK/BK
 evolution equations  \cite{jimwlk,bk}, an infinite set of coupled
 nonlinear equations for the different Wilson line correlators which
 systematically incorporate small-x gluon emission to all orders \cite{jimwlk}. In
 the large-$N_c$ limit, the  JIMWLK evolution equations reduce to the
 Balitsky-Kovchegov (BK) equation \cite{bk}, a closed-form equation for the evolution
of the dipole amplitude. The running coupling BK (rcBK) equation \cite{bk,ja-yk,bb} has 
the following simple form :
\begin{equation}
  \frac{\partial\mathcal{N}_{A(F)}(r,x)}{\partial\ln(x_0/x)}=\int d^2{\vec r_1}\
  K^{{\rm run}}({\vec r},{\vec r_1},{\vec r_2})
  \left[\mathcal{N}_{A(F)}(r_1,x)+\mathcal{N}_{A(F)}(r_2,x)
-\mathcal{N}_{A(F)}(r,x)-\mathcal{N}_{A(F)}(r_1,x)\,\mathcal{N}_{A(F)}(r_2,x)\right]\,,
\label{bk1}
\end{equation}
where the evolution kernel $K^{{\rm run}}$ using Balitsky`s
prescription \cite{bb} for the running coupling is defined as,
\begin{equation}
  K^{{\rm run}}(\vec r,\vec r_1,\vec r_2)=\frac{N_c\,\alpha_s(r^2)}{2\pi^2}
  \left[\frac{1}{r_1^2}\left(\frac{\alpha_s(r_1^2)}{\alpha_s(r_2^2)}-1\right)+
    \frac{r^2}{r_1^2\,r_2^2}+\frac{1}{r_2^2}\left(\frac{\alpha_s(r_2^2)}{\alpha_s(r_1^2)}-1\right) \right],
\label{kbal}
\end{equation}
with $\vec r_2 \equiv \vec r-\vec r_1$. For the running coupling in the above
equation we use the scheme proposed in Ref.~\cite{jav1} at one-loop
level. Notice that in the master equations (\ref{final},\ref{bk1}),
the impact-parameter dependence of the collisions was
ignored. However, for the min-bias analysis considered here
this may not be important. Nevertheless, it has been shown by
several studies that impact-dependence of the BK equation is important at
very large rapidities and for fixed centralities \cite{a00,jav0,a000,bkb,amir2}, 
though it is very challenging to implement numerically.

\section{numerical results and predictions} 
We now evaluate the single inclusive hadron production cross section numerically. To
do this, we will use the NLO MSTW 2008 PDFs \cite{mstw} and the 
NLO KKP FFs \cite{kkp}. We have checked that AKK FFs \cite{AKK} are also 
consistent with our results. We assume the factorization scale in the FFs and the
PDFs to be $Q=p_T$.

The only input for the rcBK equation is the initial conditions
for the evolution of the dipole amplitude which is commonly taken to be a  
McLerran-Venugopalan (MV) type model \cite{mv0}:
 \begin{equation}
\mathcal{N}(r,Y\!=\!0)=
1-\exp\left[-\frac{\left(r^2\,Q_{0s}^2\right)^{\gamma}}{4}\,
  \ln\left(\frac{1}{\Lambda\,r}+e\right)\right]\ ,
\label{mv}
\end{equation}
where $\Lambda=0.241$ GeV and $\gamma = 1.119$ \cite{jav0,jav1}. Notice that global fits to 
structure functions in DIS in the small-x region show that $\gamma>1$ is 
preferable and $\gamma \simeq 1.119$ provides a good fit to the DIS data \cite{jav1}. 
The onset of small-x evolution is taken at
$x_0=0.01$.  Then the only free parameter is the initial value of the
saturation scale $Q_{0s}$ (probed by quarks) for proton and nucleus at
$x_0= 0.01$. The value of $Q_{0s}$ can be also fixed via a fit to the
structure functions in electron-proton and electron-nucleus scatterings in the small-x region. 
In order to investigate the sensitivity
of our results to the choice of the parameter sets, we take $Q_{0s}$
as a free parameter and let it be determined by the RHIC data in proton-proton (pp)
and deuteron-gold (dAu) collisions. In this way, we will also examine whether a
universal description of various low-x data can be achieved by the 
rcBK evolution equation. 

In order to facilitate a comparison of the solution to rcBK dipole evolution 
equation, we also employ the DHJ dipole parametrization~\cite{dhj2} 
for $N_{A(F)}$ which has been used to describe the forward
rapidity data in dAu collisions at RHIC. We refer the reader to
Ref.~\cite{dhj2} for the details of the DHJ dipole model. The dipole
scattering amplitude in the DHJ model is simply given by 
\begin{equation}
\mathcal{N}_{A}^{\text{DHJ}}(k_T,x)= \int d^2\vec r e^{-i\vec k_T.\vec r}\left(1-
\exp\left[-\frac{1}{4}\left(r^2\,Q_{s}^2(x)\right)^{\gamma^{\text{DHJ}}(\mathcal{Q}_T,x)}\right]
\right)\ ,
\label{dhj-d}
\end{equation}
where the anomalous dimension $\gamma^{\text{DHJ}}$ in the DHJ model is 
parameterized as,
\begin{equation}
\gamma^{\text{DHJ}}(\mathcal{Q}_T,x)=\gamma_s+(1-\gamma_s)\frac{\log\left(\mathcal{Q}_T^2/Q_s^2(x)
\right)}{\lambda y+d\sqrt{y}+\log\left(\mathcal{Q}_T^2/Q_s^2(x)\right)},
\label{dhj-g}
\end{equation}
with $y=\log(1/x)$ and the scale $\mathcal{Q}_T$ in the anomalous dimension
is related to the inverse transverse size of the dipole $\mathcal{Q}^2_T\approx
1/r^2$. Saturation scale $Q_s(x)$ is defined as
$Q^2_s(x)=A^{1/3}_{eff}(x_{s0}/x)^{\lambda}$ with $A_{eff}=18.5$ for
the minimum bias dAu collisions. Here we are interested in
investigating whether the new inelastic contribution, second line in
Eq.~(\ref{final}), will affect the description of the RHIC data
\cite{dhj2}. To this end, we take the same parameters for the DHJ
dipole parametrization as employed in Ref.~\cite{dhj2} which provides
a good description of the RHIC data without the presence of the
inelastic contribution namely $\alpha_s=0$. The parameters
$\lambda=0.3$ and $x_{s0}=10^{-4}$ were extracted from a fit to HERA
data, and parameter $d$ was fitted to the RHIC data and set to $d =
1.2$ \cite{dhj2}. The anomalous dimension in \eq{dhj-g} runs from 
the LO BFKL value $\gamma_s=0.628$ at small $x$ to the DGLAP 
value $\gamma^{\text{DHJ}}\to 1$. This model incorporates  
the geometric scaling window as expected from
the BK equation \cite{gs} consistently.

 Here our aim is not to fit the data
but to use the best theoretical tools available in low-x physics to 
highlight the uncertainties involved in making robust predictions
from the CGC formalism for the upcoming proton-nucleus collisions at the LHC.  
Therefore, we take $K=1$ throughout this paper.

\begin{figure}[t]
              \includegraphics[width=8 cm] {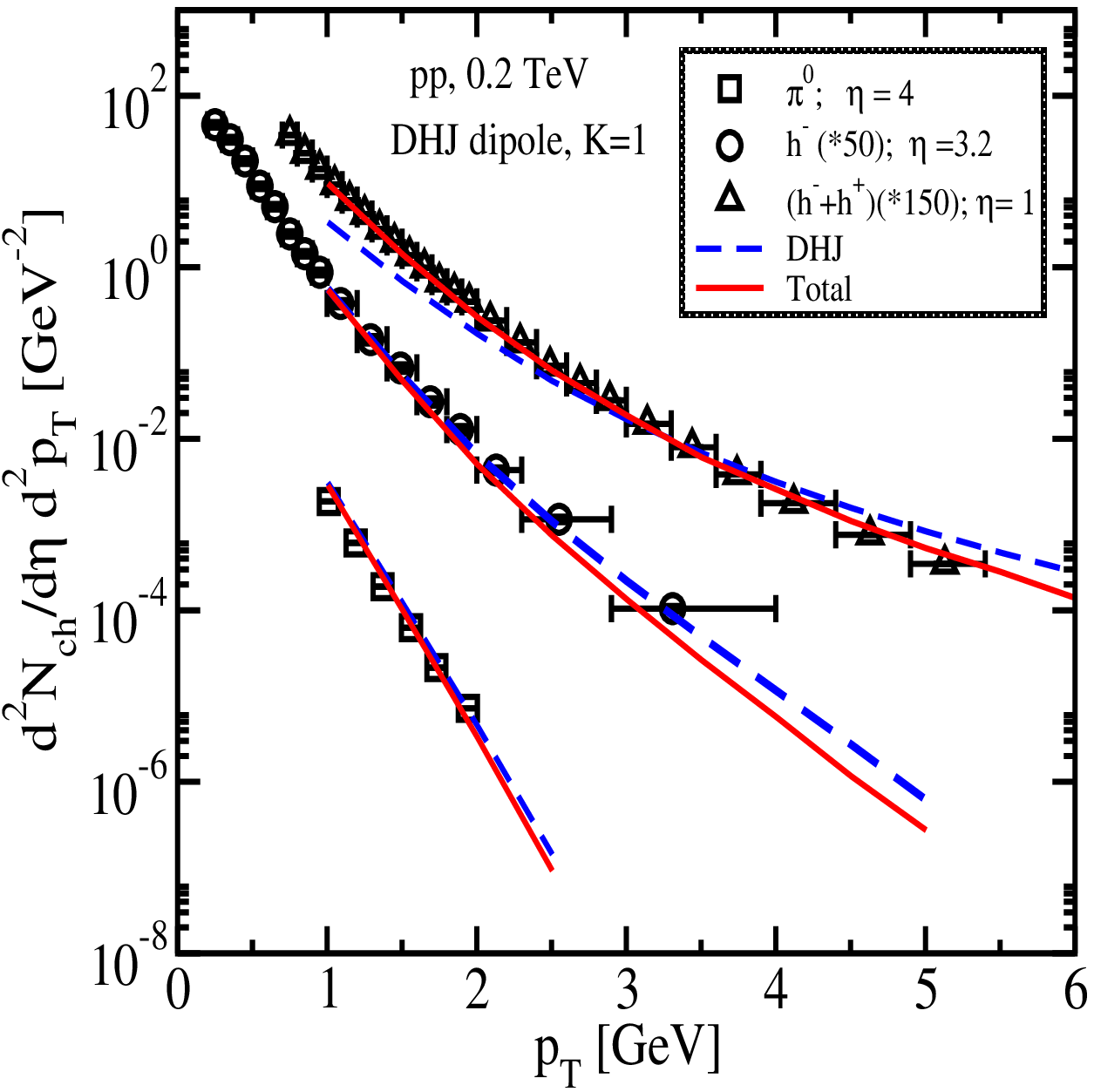}
              \includegraphics[width=8 cm] {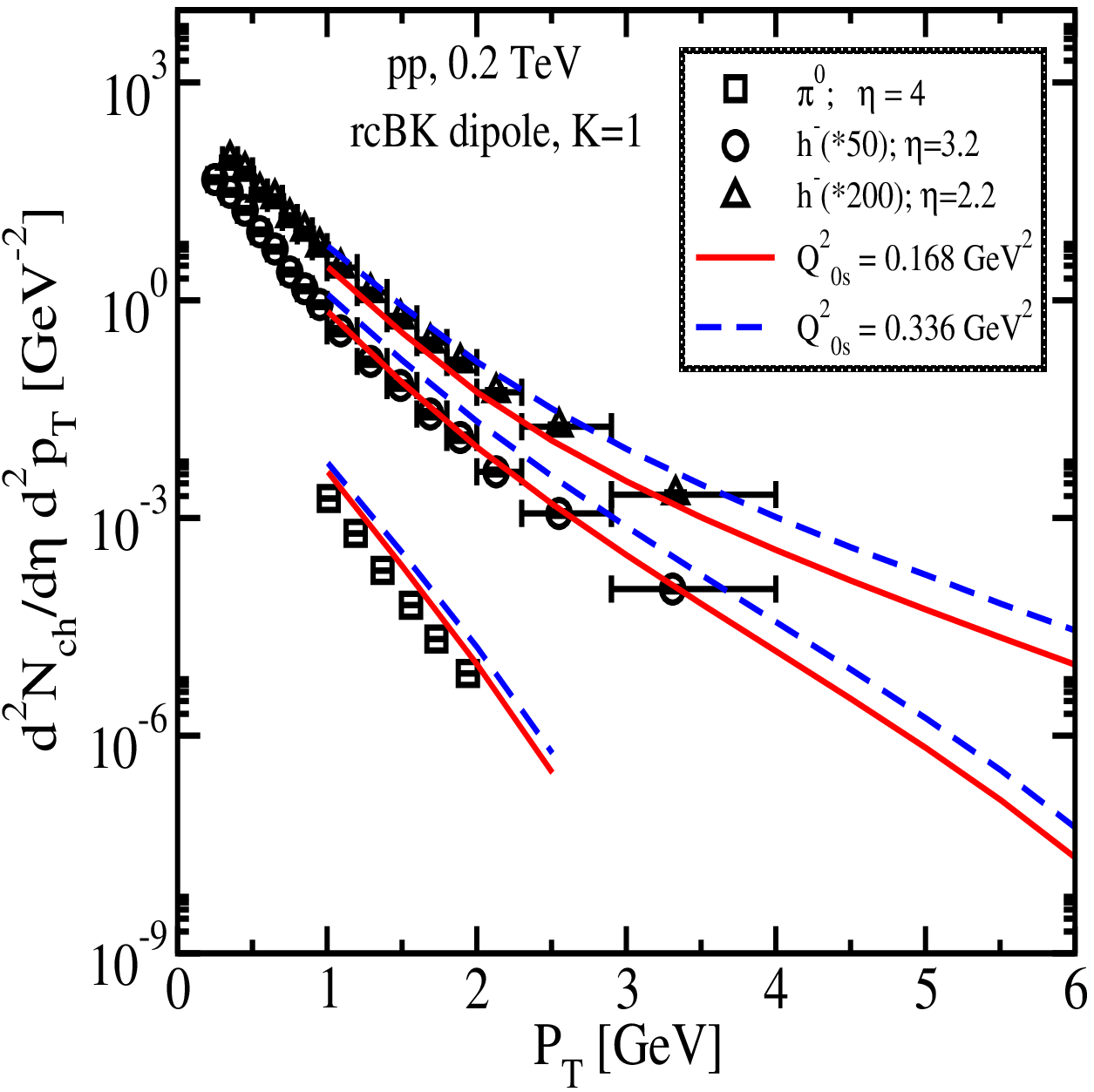}
              \includegraphics[width=8 cm] {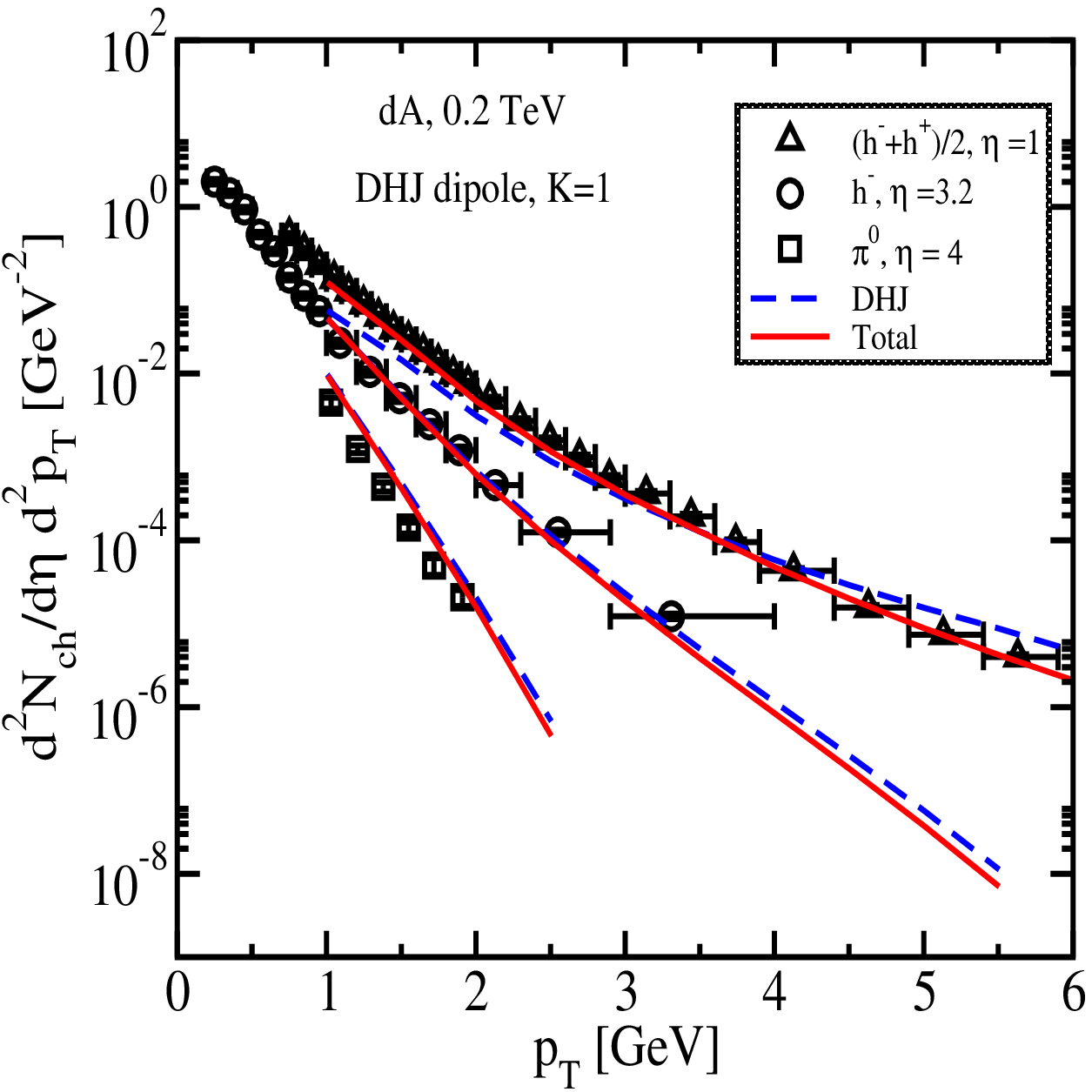}
              \includegraphics[width=8 cm] {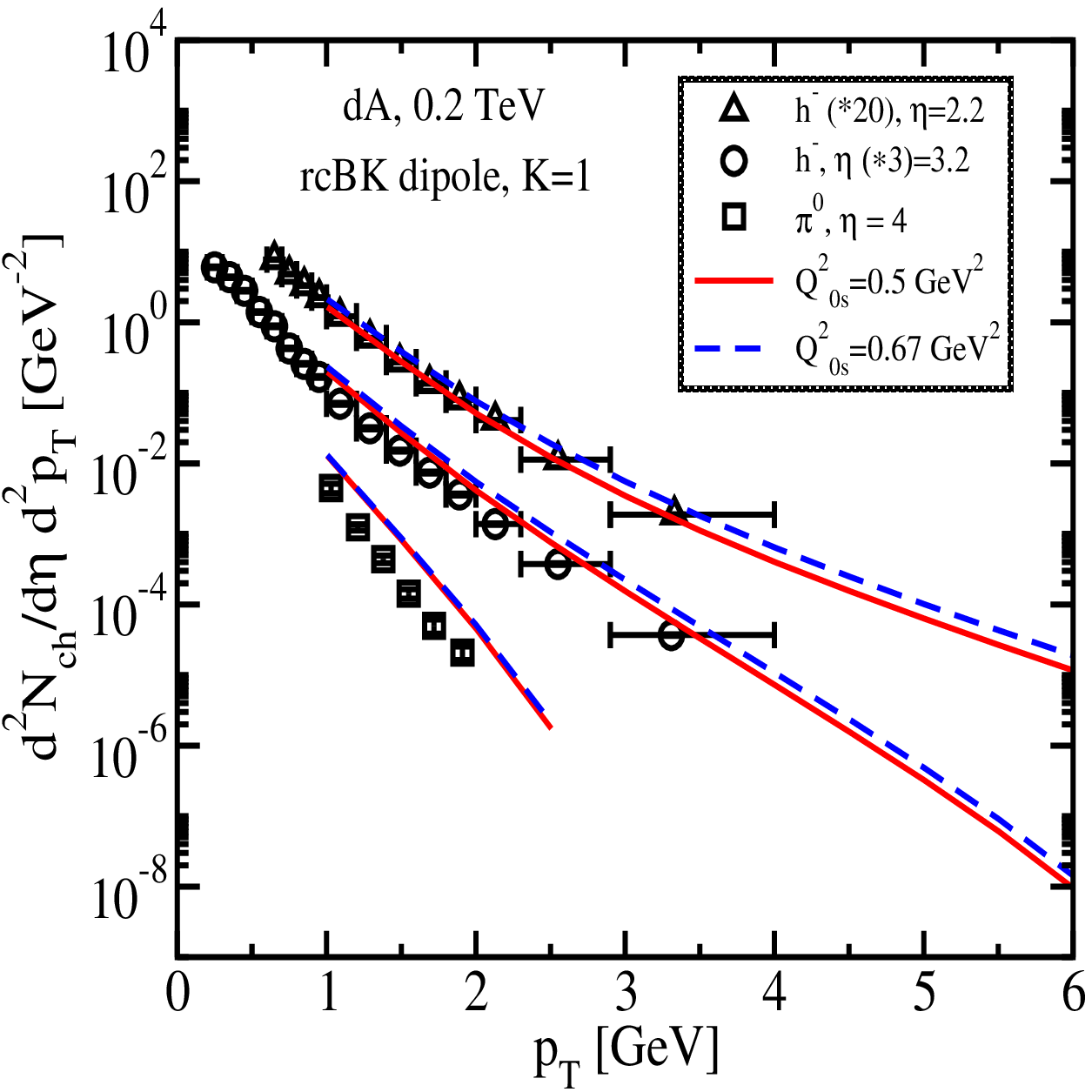}
\caption{Single inclusive hadron production in proton-proton (upper panel)
 and deuteron-gold (lower panel) collisions at different
 pseudo-rapidities at RHIC obtained by the solution of the
 running-coupling BK equation, the so-called rcBK (right) and the DHJ
 (left) dipole model.  Right: dashed and full lines refer to the
 results coming from the rcBK equation corresponding to two different
 initial values for the saturation scale at $x_0=0.01$. We
 have taken $\alpha_s=0.1$ in \eq{final} for all curves.  
Left: dashed and full lines
 refers to the results when $\alpha_s=0$ (the DHJ term or 
the elastic contribution) and
 $\alpha_s=0.1$ (for the inelastic term), respectively. We have taken $K=1$ 
 in all panels. The experimental data are from Ref.~\cite{exp}. }
\label{fig-pp1}
\end{figure}

In \fig{fig-pp1} we show the single inclusive hadron production yields
in pp and dAu collisions at RHIC $\sqrt{s}=0.2$ TeV at different
rapidities using the DHJ parametrization of the dipole cross section
as well as the rcBK dipole solution.  In order to investigate the
contribution of the inelastic term to single inclusive hadron production, we
also show the results without this term, namely $\alpha_s=0$ in
\eq{final}, denoted as DHJ in \fig{fig-pp1} (left). The inelastic
contribution term in \eq{final} is explicitly proportional to
$\alpha_s$. Notice that in the derivation of this formula at the
leading-twist order, $\alpha_s$ was assumed to be a fixed
parameter. It is not a priori obvious whether the running-coupling
corrections to \eq{final} can be simply incorporated by replacing
$\alpha_s$ to a running $\alpha_s(Q)$. It has been shown for example
that in $k_T$-factorization formulation, the running coupling effect
changes the equation \cite{kov}. We have checked that $\alpha_s\approx
0.05\div 0.15$ in \eq{final} gives a reasonable description of RHIC
data for both pp and dAu collisions. It is clear that inclusion of the
inelastic terms improves the $p_T$ dependence of the cross section
closer to mid-rapidity while there is no visible contribution at the
most forward rapidity considered. This is more clearly seen in
\fig{fig-pp1} at the upper-left panel, where the inelastic contribution
is seen to significantly improve the description of the data for more central
collisions at $\eta=1$ and makes the $p_T$-spectra steeper in
agreement with the data. For more forward collisions, the available
phase space is limited and inelastic contributions are consequently 
negligible independent of the value of the strong coupling $\alpha_s$.

As we have already pointed out, the value of $Q_{0s}$ at
$x_0=0.01$ is the only free-parameter left to be
fixed for a given solution of the rcBK equation. In
\fig{fig-pp1} (right), we also show the effect of various choices 
for the initial saturation scale $Q_{0s}$. In the case of pp RHIC data,
we found that values in $Q_{0s}^2= 0.168\div 0.336~\text{GeV}^2$ range 
give a consistent description of data. However, a smaller
value of $Q_{0s}^2=0.168~\text{GeV}^2$ may be more preferable,
specially at very forward rapidities. This is understandable
since the available phase space for multiple re-scattering is 
limited at very forward rapidity. Therefore, a lower initial saturation scale
is required to describe the cross-section. We note that HERA data on proton
structure functions prefer a lower value for the proton
initial saturation scale\footnote{The parameter 
set $Q_{0s}^2= 0.168~\text{GeV}^2$
and $\gamma=1.119$ that we used in \fig{fig-pp1} also gives excellent
description of the structure function data in e+p collisions with
$\chi^2/d.o.f.=1.104$ \cite{jav1}.} $Q_{0s}^2= 0.168~\text{GeV}^2$\cite{jav1}.  
In the case of mini-bias dAu collisions, the
initial nuclear (gold) saturation scale within $Q_{0s}^2=0.5 \div
0.67~\text{GeV}^2$ is consistent with the RHIC data within the error
bars. Unfortunately the available DIS data for nuclear targets are  
limited and have large experimental uncertainties. It is therefore 
difficult to pin down the exact value of $Q_{0s}$ for nuclei based on
only DIS data, see also Ref.~\cite{raju}. In the case of proton-nucleus collisions, due to 
theoretical uncertainties and rather large experimental data errors, 
it is also not possible to uniquely fix the initial value of $Q_{0s}$.
Nevertheless, the extracted values of initial nuclear saturation scale
here are compatible with values extracted in other studies, see Refs.~\cite{ja-cm,jav0} and
reference therein.

It is seen from \fig{fig-pp1} that for the description of $\pi^0$
production in both pp and dAu collisions at very forward rapidity
($\eta=4$) at RHIC, a $K$-factor of $\sim 0.4\div 0.6$ may be needed. The
necessity of such a small $K$-factor at very forward rapidity at RHIC
$\sqrt{s}=200$ GeV was also shown in Ref.~\cite{ja-cm} where the
inelastic contribution was ignored. Notice that as we have already pointed
out, the effect of inelastic contribution at very forward rapidities at
RHIC energy $\sqrt{s}=200$ GeV is negligible numerically, see \fig{fig-pp1}. 
It is possible that at very forward rapidities at RHIC energy, other mechanisms 
also partially contribute to the hadron production, see for
 example Refs.~\cite{boris,mark}. 

Next, we present our predictions for single inclusive hadron
production at the LHC in terms of the nuclear modification 
factor $R_{pA}$ hoping that
some of the theoretical uncertainties, such as sensitivity to $K$ factors, will be reduced. 
The Nuclear modification factor $R_{pA}$ is defined as
\begin{equation}
R_{pA}=\frac{1}{N_{coll}}\frac{dN^{p A \rightarrow h X}}{d^2p_T d\eta}/
\frac{dN^{p p \rightarrow h X}}{d^2p_T d\eta},
\end{equation}
where $N_{coll}$ is the number of binary proton-nucleus collisions. We
take $N_{coll}=6.5, 7.4$ at $\sqrt{s}=4.4$ and $8.8$ TeV,
respectively\footnote{In order to compare our $R_{pA}$ predictions with experimental data,
one may need to rescale $R_{pA}$ by matching the normalization $N_{coll}$
to the experimental value.} \cite{ncoll}. 
\begin{figure}[t]
               \includegraphics[width=8 cm] {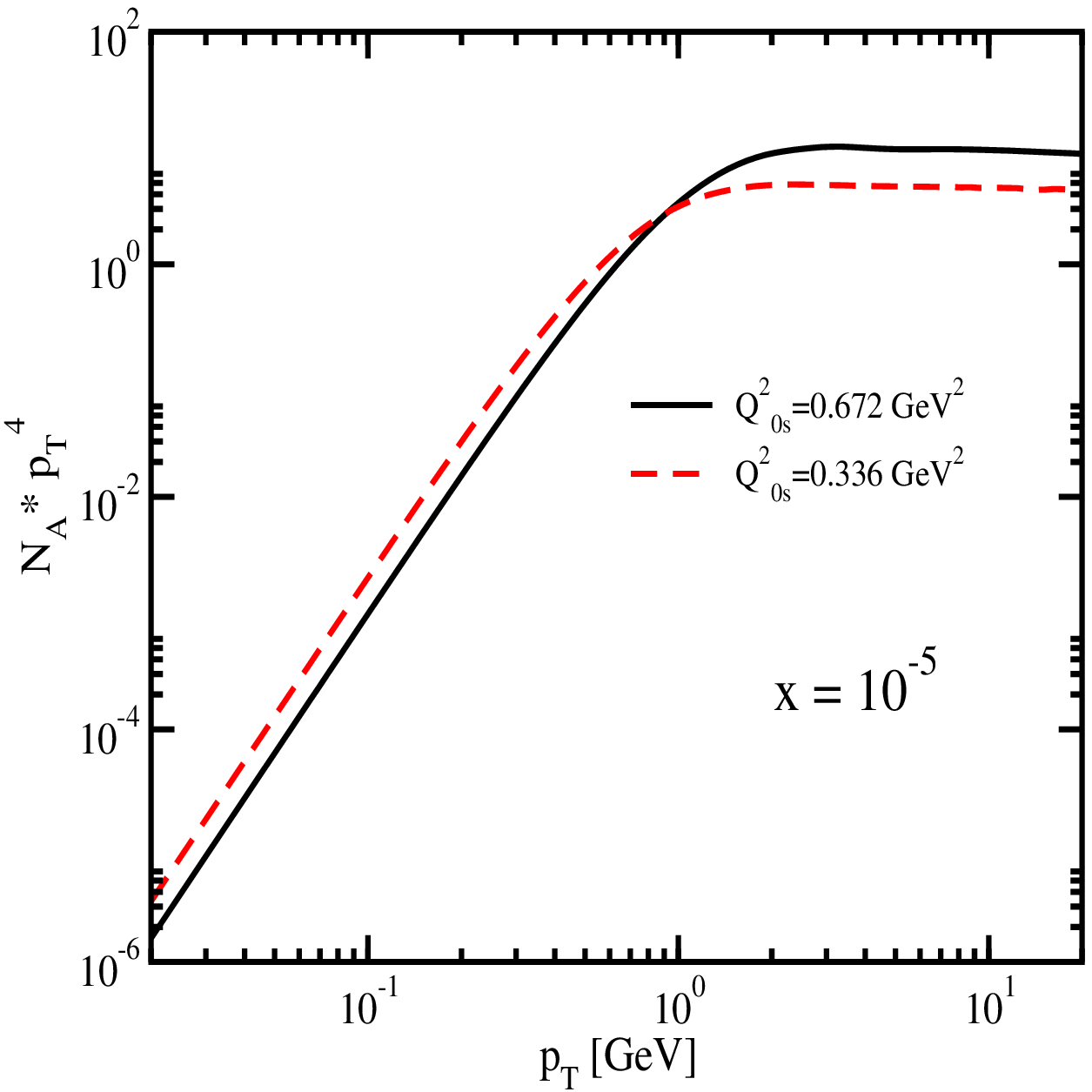}
               \includegraphics[width=8 cm] {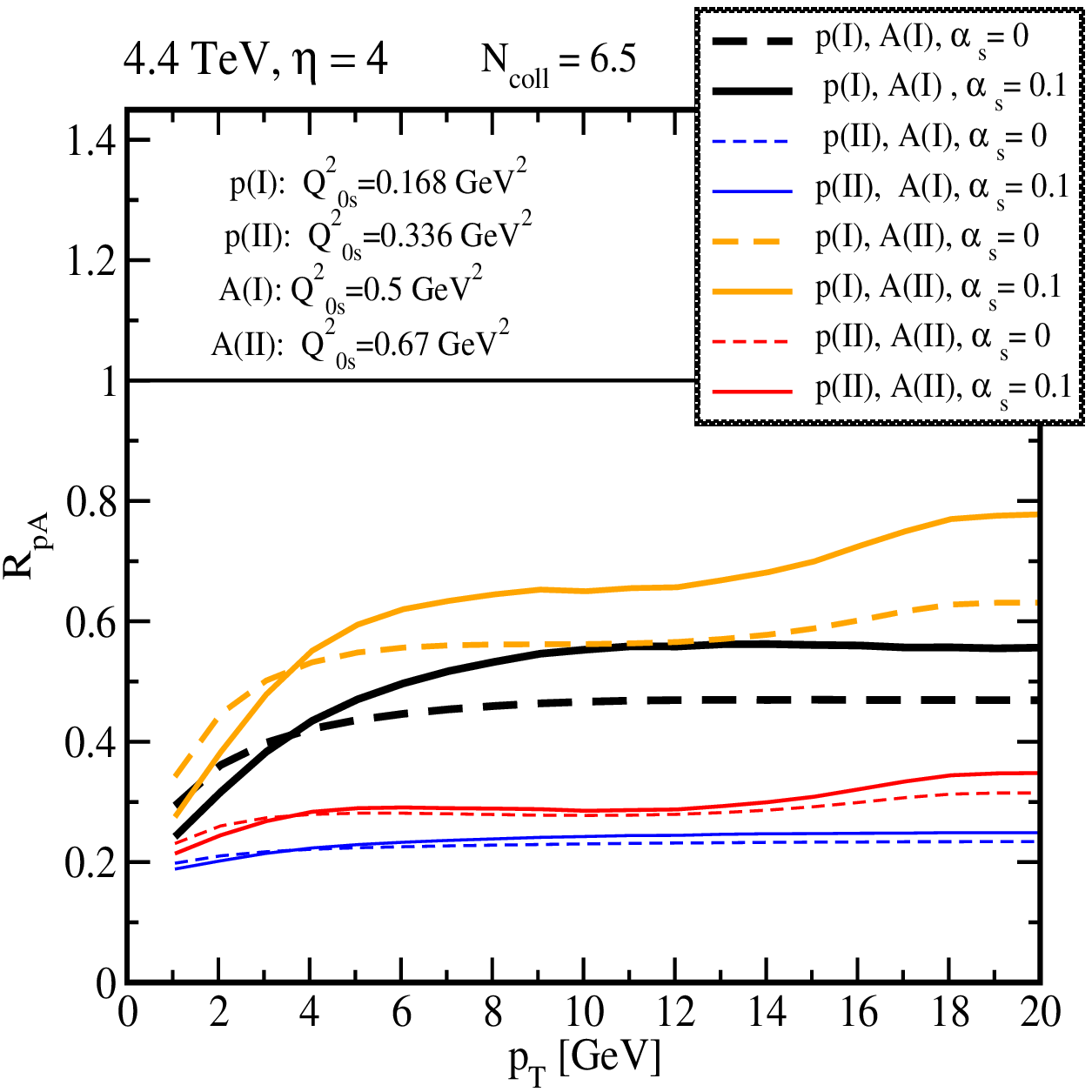}
               \caption{Right: nuclear modification factor
               $R_{pA}$ for inclusive charged hadrons $h^{+}+h^{-}$ 
production at the LHC $\sqrt{s}=4.4$ TeV and $\eta=4$
               coming from the solutions of the rcBK with different
               initial values for the saturation scale (at $x_0=0.01$)
               for proton and nucleus. The dashed and full lines refer
               to the cases when the cross-section in both pp and
               pA collisions was obtained via Eq.~(\ref{final}) by
               taking $\alpha_s=0$ (only elastic contribution) and
               $\alpha_s=0.1$ respectively. Left: the scaled unintegrated
               gluon distribution $N_A(x,p_T) \times p_T^4$ as a
               function of transverse momentum $p_T$ at a fixed $x=10^{-5}$ 
obtained from the
               rcBK equation with two different initial values for the
               saturation scale $Q_{0s} (x_0=0.01)$. }
\label{fig-lhc1}
\end{figure}

\begin{figure}[t]                            
                \includegraphics[width=8 cm] {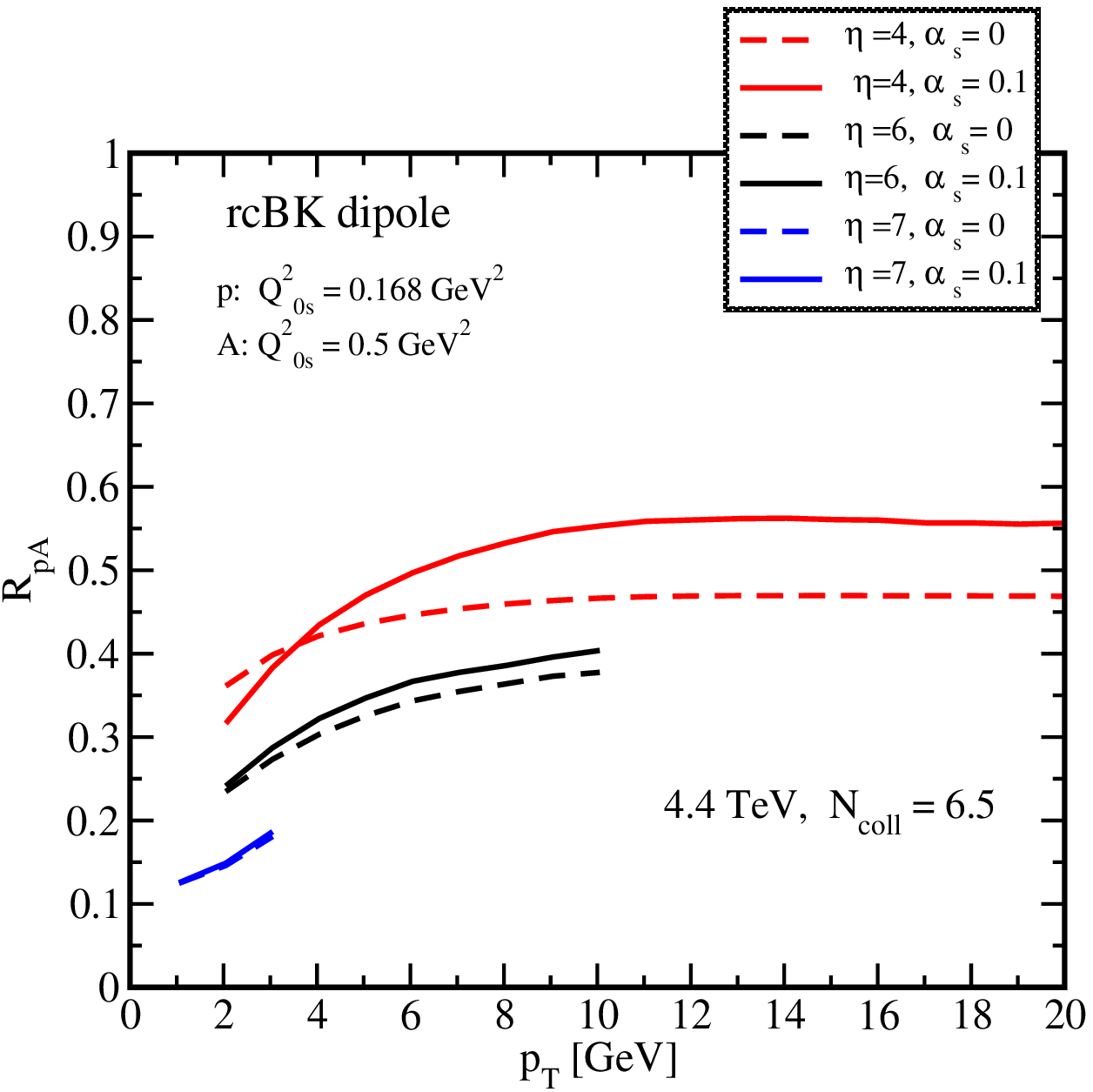}
                 \includegraphics[width=8 cm] {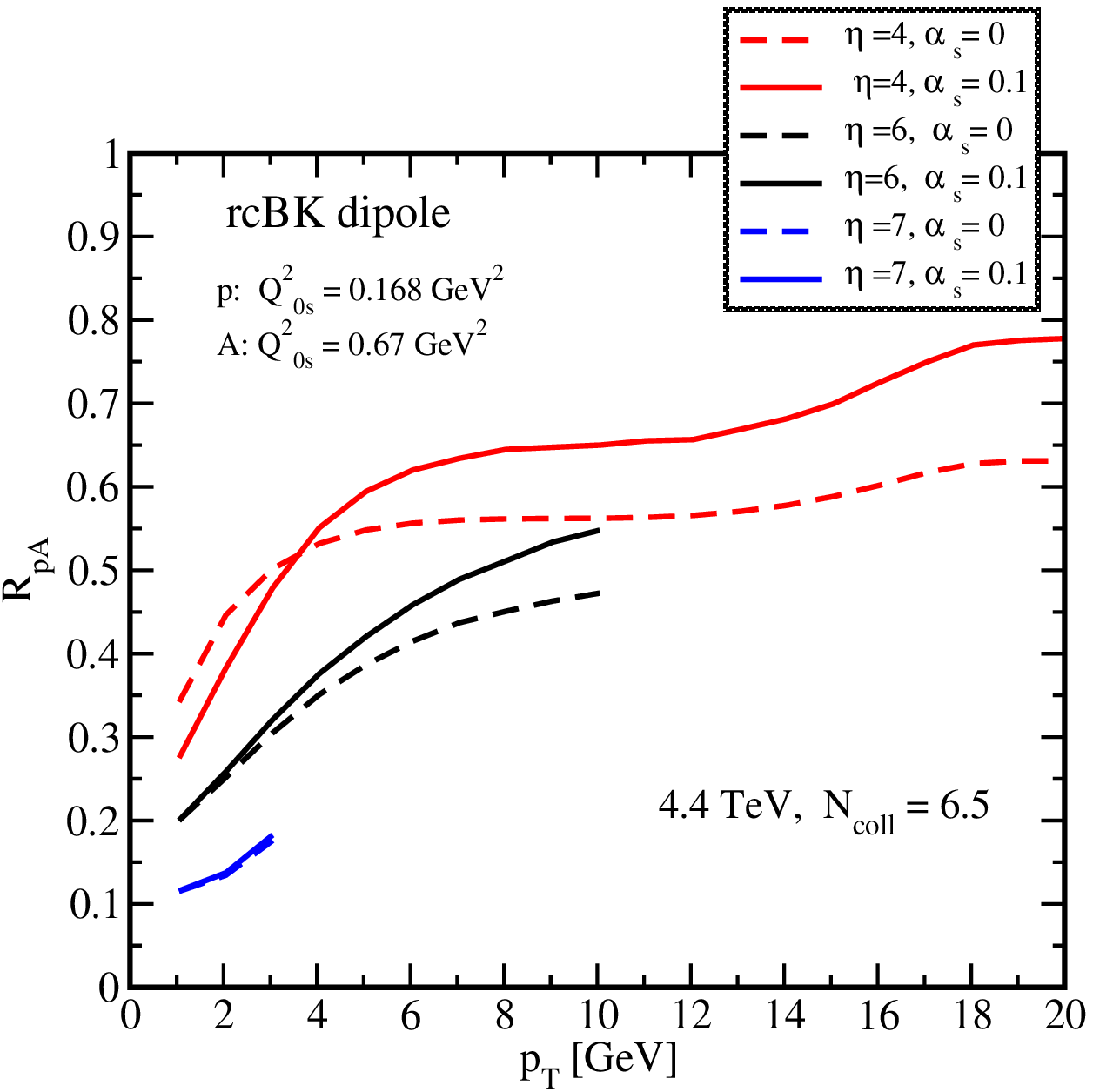}
\caption{Nuclear modification factor
               $R_{pA}$ for $h^{+}+h^{-}$ production in proton-nucleus
               collisions at the LHC ($\sqrt{s}=4.4$ TeV) at different
               rapidities (from top to bottom: $\eta=4,6, 7$) obtained
               from the solution of rcBK equation assuming two different initial nuclear saturation
               scales $Q_{0s}^2=0.5~\text{GeV}^2$ (left) and
               $Q_{0s}^2=0.67~\text{GeV}^2$ (right). In both panels the
               initial saturation scale for proton was taken
               $Q_{0s}^2=0.168~\text{GeV}^2$. The effect of different value for the strong coupling $\alpha_s$ in \eq{final} is also shown. }
\label{fig-lhc2}
\end{figure}

In \fig{fig-lhc1} (right), we show the nuclear modification factor
$R_{pA}$ for inclusive charged hadrons $h^{+}+h^{-}$ production 
at $\sqrt{s}=4.4$ TeV and $\eta=4$ obtained from different
solutions of the rcBK equation corresponding to different values of 
$Q_{0s} (x_0=0.01)$ extracted from RHIC data (see description of
\fig{fig-pp1}). We also show the contribution of the inelastic term 
by showing the results due to only
the DHJ term ($\alpha_s=0$). The value of the strong-coupling in the 
inelastic term in \eq{final}
is set to $\alpha_s=0.1$ (the same value was taken in \fig{fig-pp1}). 
It is obvious that taking
different values for the saturation scale $Q_{0s} (x_0=0.01)$ for
proton and nuclear targets significantly changes the nuclear
modification factor. Therefore, the measurement of $R_{pA}$ provides
vital information about the initial saturation scale of target and
small-x evolution dynamics. Inclusion of the inelastic term changes 
$R_{pA}$ and makes it increase faster at high-$p_T$, see also \fig{fig-lhc2}. 
Notice that rcBK solutions taken here approximately reproduce the 
perturbative power-law behavior
of the dipole-amplitude $N_{A(F)}\sim 1/p_T^4$ at high-$p_T$, see
\fig{fig-lhc1} (left). We recall that the parameters of rcBK solutions 
used here  
were obtained from a fit to HERA data for virtuality 
$Q^2\in [0.25, 45]~\text{GeV}^2$ \cite{jav1}. 
Therefore, our results at very high-$p_T$ may be less reliable. 

\begin{figure}[t]                            
                \includegraphics[width=8 cm] {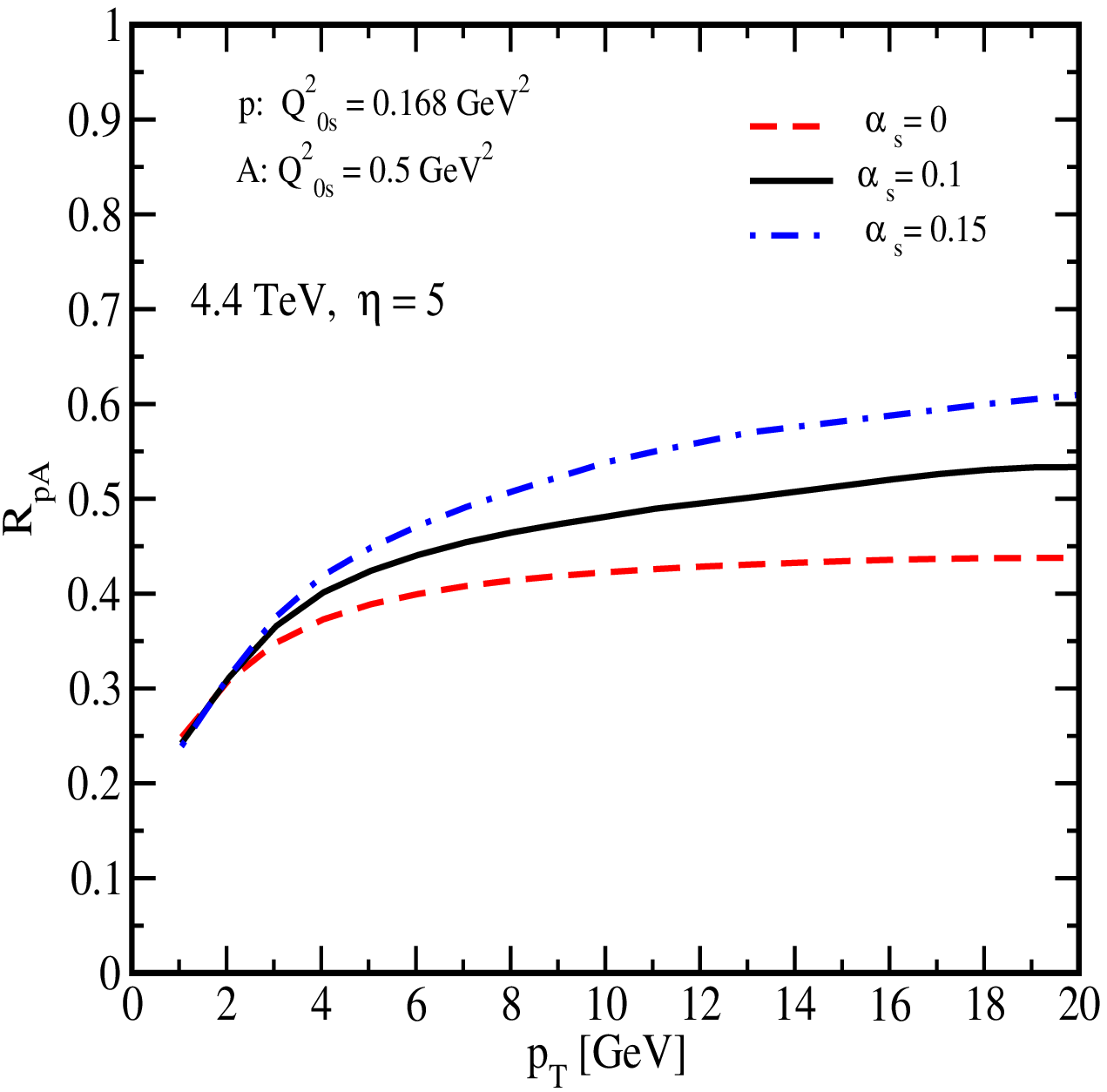}
                \includegraphics[width=8 cm] {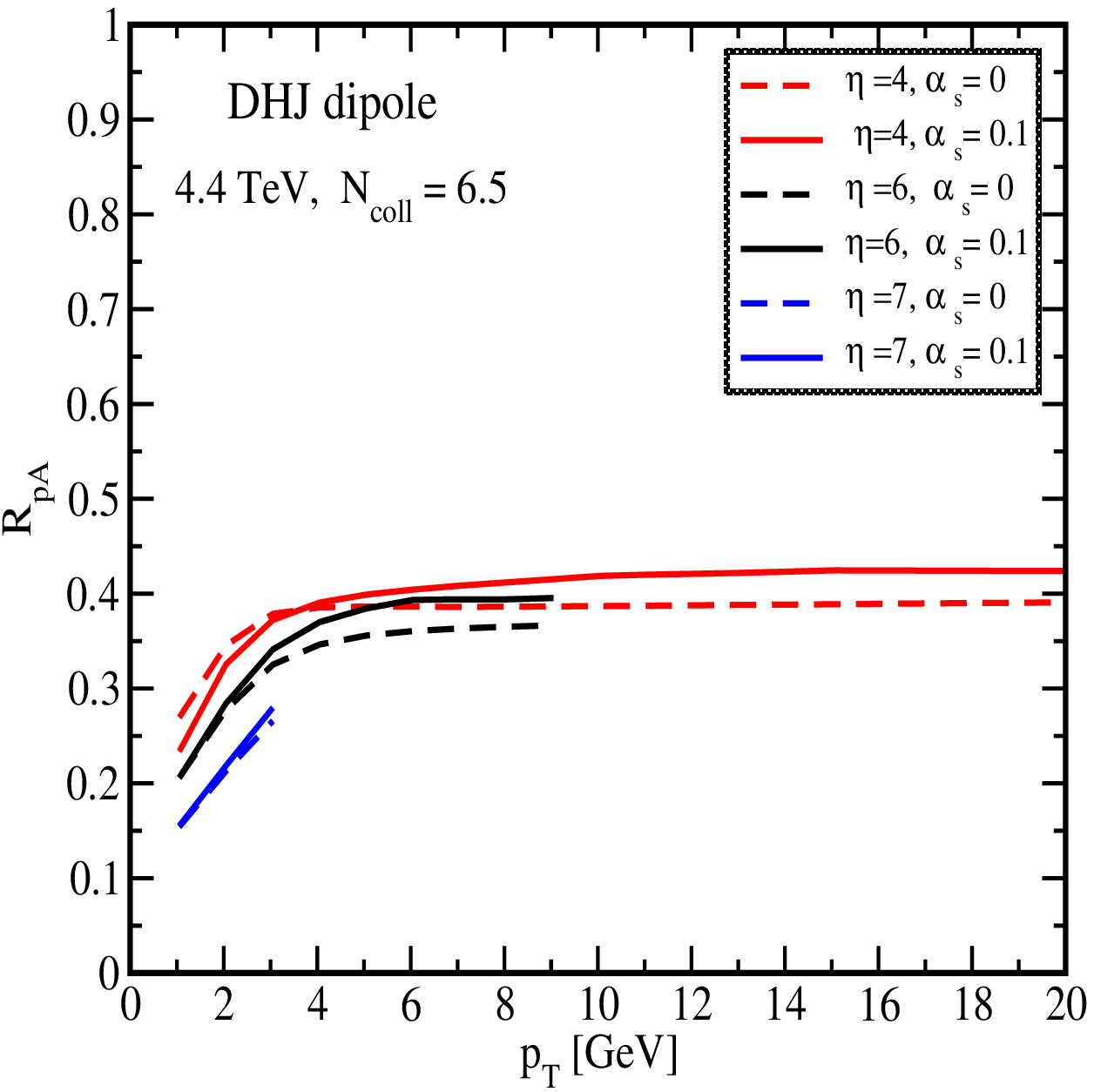}
                 \caption{Right:  
$R_{pA}$ for inclusive charged hadron production at different rapidities at the LHC obtained from the DHJ parameterization of the dipole profile \eq{dhj-d}, 
with different values for the strong coupling in \eq{final}. $R_{pA}$ for inclusive charged hadron production for various values of the 
strong coupling constant $\alpha_s$ in \eq{final} 
at the LHC ($\sqrt{s}=4.4$ TeV and $\eta=5$) obtained by the rcBK equation \eq{bk1}. }
\label{fig-lhc3}
\end{figure}

In \fig{fig-lhc2}, we show our predictions for $R_{pA}$ for
$h^{+}+h^{-}$ production at $\sqrt{s}=4.4$ TeV and $\eta=4-7$ using
the solution of the rcBK dipole evolution equation, \eq{bk1}, assuming
initial nuclear saturation scales of $Q_{0s}^2=0.67~\text{GeV}^2$
(right panel) and $Q_{0s}^2=0.5~\text{GeV}^2$ (left panel). In both
panels we have assumed the initial saturation scale of proton
to be $Q_{0s}^2=0.168~\text{GeV}^2$. We note that a larger
initial saturation scale for the  nucleus leads to a faster rise of $R_{pA}$
with transverse momentum . For comparison, in \fig{fig-lhc3} (right
panel) we show the corresponding $R_{pA}$ obtained using the DHJ dipole model,
defined in \eq{dhj-d}. It is seen that both approaches lead to a suppression 
of $R_{pA}$ at forward rapidities at the LHC and that the DHJ parameterization
leads to a flatter transverse momentum dependence. We recall that both the 
rcBK solution and the DHJ model provide a reasonable description of RHIC data.

In order to highlight the uncertainties associated with the different choices of the strong-coupling 
constant in \eq{final} more clearly, in \fig{fig-lhc3} (left) we show $R_{pA}$ for three different 
values of $\alpha_s$ namely, $\alpha_s=0$ corresponding to the elastic term only, and 
$\alpha_s=0.1,0.15$ for the inelastic contribution.  It is seen from 
Figs.~\ref{fig-lhc1}, \ref{fig-lhc2}, \ref{fig-lhc3} that at rapidities close to mid-rapidity, 
increasing $\alpha_s$ reduces $R_{pA}$ while at very forward rapidities and high-$p_T$ the opposite happens. 

\begin{figure}[bht]                            
                                  \includegraphics[width=8 cm] {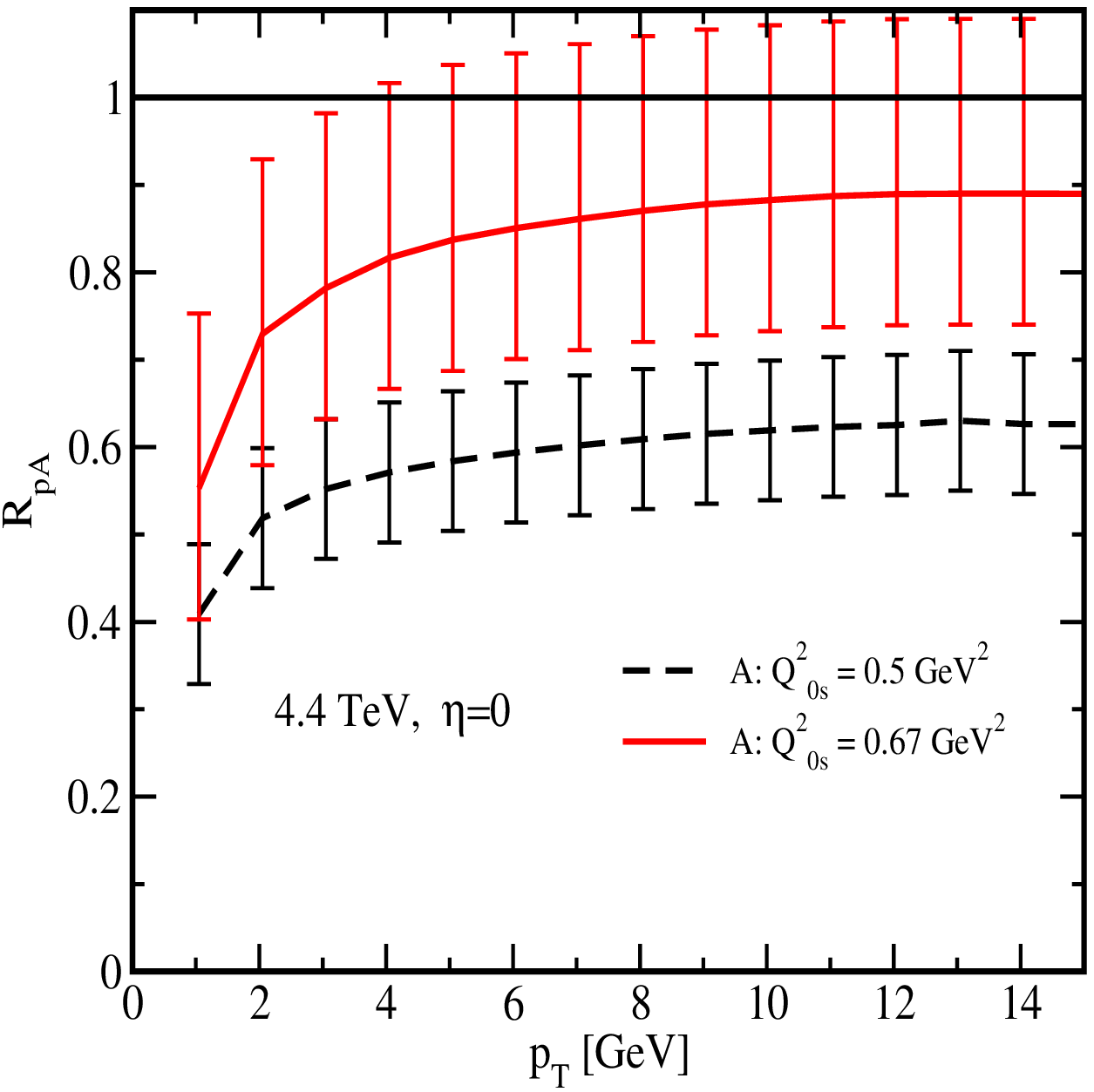}
                 \includegraphics[width=8 cm] {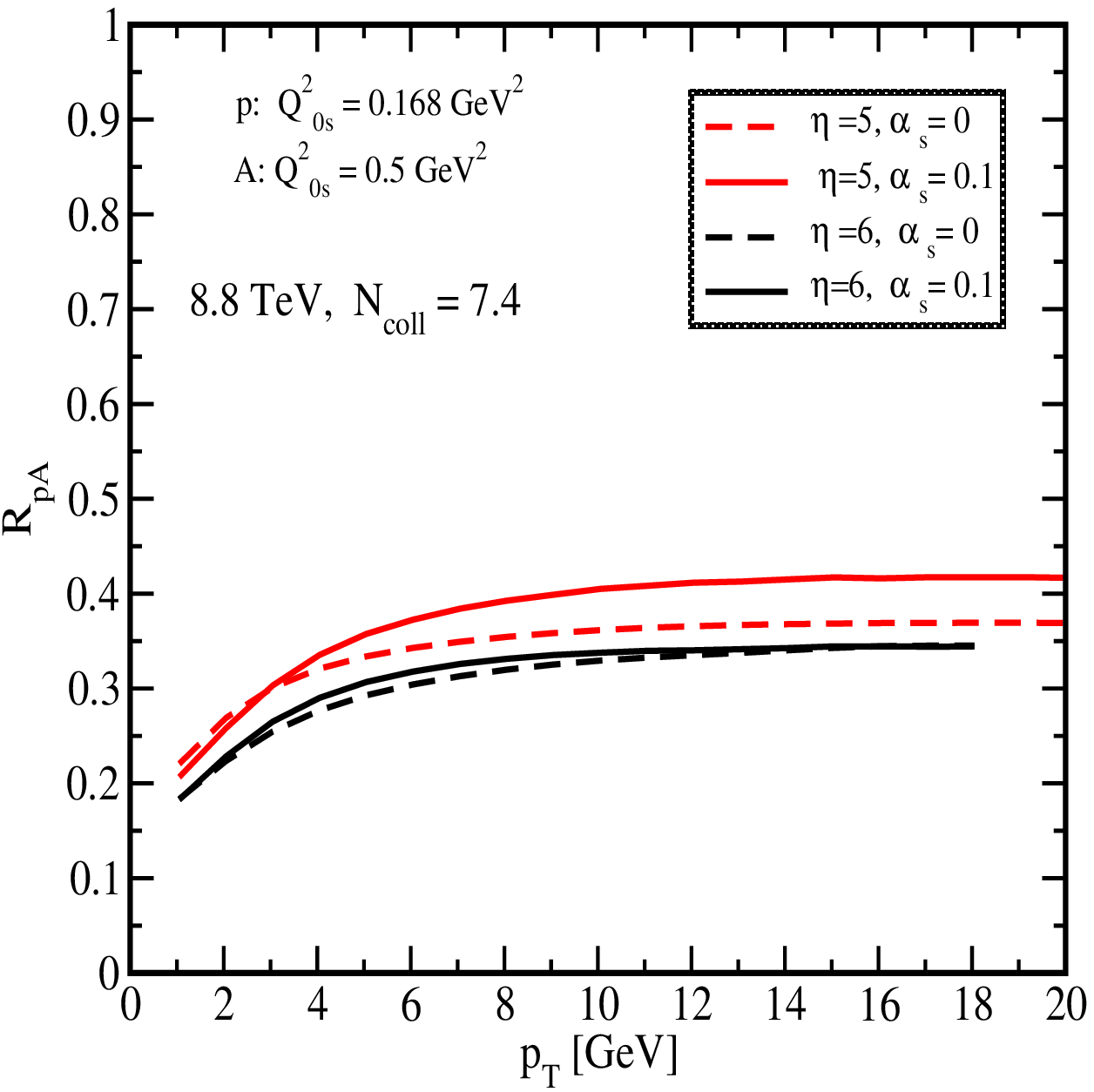}
                 \caption{Right: nuclear 
modification factor $R_{pA}$ for $h^{+}+h^{-}$ production in proton-nucleus 
collisions at the LHC ($\sqrt{s}=8.8$ TeV) at different rapidities 
obtained from the solution of rcBK dipole evolution equation 
(\ref{bk1}) with different values for the strong coupling constant 
in \eq{final}. Left:  $R_{pA}$ for
                 $h^{+}+h^{-}$ production in proton-nucleus collisions 
at the LHC in midrapidity ($\sqrt{s}=4.4$ TeV, $\eta=0$) for two different initial nuclear
saturation scales of $Q_{0s}^2 = 0.5, 0.67\, \text{GeV}^2$ extracted from
RHIC data. The initial saturation scale for proton is taken to be $Q_{0s}^2  = 0.168~\text{GeV}^2$. The
theoretical error bars mainly show the uncertainties associated with
the choice of $\alpha_s$.}
\label{fig-lhc4}
\end{figure}

In \fig{fig-lhc4} (right), we show our predictions for $R_{pA}$ for inclusive charged
hadron production at $\sqrt{s}=8.8$ TeV and at different rapidities
obtained from the rcBK equation (\ref{bk1}) with different values of
the strong coupling in the master equation (\ref{final}). It is seen
that the energy-dependence of $R_{pA}$ from $\sqrt{s}=4.4$ to $8.8$
TeV is rather weak\footnote{Note that our results for $R_{pA}$ at the
LHC without inclusion of the inelastic term is different from
Ref.~\cite{ja-cm} mainly due to the fact that we have used a different
parameter set for the rcBK solution with $\gamma=1.119$ which was
recently suggested in Refs.~\cite{jav0,jav1}. For the sensitivity of
$R_{pA}$ to the various allowed solutions of the rcBK equation, see
\fig{fig-lhc1} and related discussions.}. From
Figs.~\ref{fig-lhc2}, \ref{fig-lhc3}, we also note that at very forward
rapidities the uncertainty associated with the choice of $\alpha_s$ is
reduced. This is in accordance with the fact that the effect of
inelastic contribution at very forward rapidities is negligible, see
also \fig{fig-pp1}. 

Our prediction for $\eta=0$ at the LHC energy $4.4$ TeV  is shown in \fig{fig-lhc4} (left), 
using the solution to the rcBK evolution equation and assuming two different initial nuclear
saturation scales of $Q_{0s}^2 = 0.5, 0.67\, \text{GeV}^2$ (extracted from
RHIC data). In \fig{fig-lhc4}, we assumed the initial saturation scale for proton to be
$Q_{0s}^2 = 0.168~\text{GeV}^2$ (extracted from RHIC and HERA data). The
theoretical error bars in \fig{fig-lhc4} show the uncertainties mainly associated with
the choice of $\alpha_s$ in \eq{final}. The observed suppression of $R_{pA}$ at
midrapidity and high $p_T$ for the case of lower initial nuclear
saturation scale $Q_{0s}^2 = 0.5\, \text{GeV}^2$ is larger compared to the
results obtained within the gluon saturation approach with
quasi-classical (Glauber) approximation \cite{AS}. We note that there
are large uncertainties in $R_{pA}$ in midrapidity at the LHC due to
the choice of the initial saturation scale for the rcBK evolution
equation (\ref{bk1}), and the value of strong-coupling constant in
\eq{final}. More importantly, large sensitivity of $R_{pA}$ to the
value of $\alpha_s$ in \eq{final} in midrapidity at the LHC indicates
that higher order corrections in
\eq{final} should be important in midrapidity at the LHC
energy. Therefore, we believe that our predictions for
$R_{pA}$ at midrapidity may be less reliable compared to our 
results for the very forward rapidity collisions.

It should be noted that the particle production cross-section given by
\eq{final} is intrinsically asymmetric, namely it treats the
projectile proton approximately in the collinear factorization
framework while treating the target proton (or nucleus) in the CGC
framework. Strictly speaking, this may be justified only for particle
production in the collision of a dilute system on a dense system, such
as particle production in mid or forward rapidity in pA collisions or
in particle production in the very forward rapidity region in
symmetric collisions, such as proton-proton or nucleus-nucleus
collisions. Therefore, our formalism can not be reliable for particle
production in midrapidity in proton-proton collisions. Unfortunately,
this is also what one needs in order to calculate the nuclear
modification factor $R_{pA}$ in midrapidity. A better approach to
particle production in midrapidity in symmetric collisions where both projectile
and target are dilute (for $p_t >> Q_s$) might be to use the $k_t$ factorization 
formalism,  proven to LO accuracy for single inclusive hadron 
production \cite{kt-p}. This is beyond the scope of the present work and we leave it for a
future study.

\section{Summary} 
We have quantitatively studied, for the first time, the contribution of both elastic and
inelastic processes to single inclusive hadron production cross section at RHIC
and the LHC using the CGC formalism. We observe that inelastic contributions to single
inclusive hadron production are significant at high transverse momentum and close to 
mid-rapidity. On the other hand, their contribution is very small in the forward rapidity
region. Furthermore, we note that inclusion of these inelastic terms makes the nuclear 
modification factor $R_{pA}$ grow faster with increasing transverse momentum. 
We make detailed predictions for $R_{pA}$ at the LHC using the numerical solution of 
the running-coupling BK equation. We have studied various theoretical uncertainties 
associated with the choice of the initial saturation scale $Q_{0s}^2$ for a proton and nucleus.
We have shown that the nuclear modification factor $R_{pA}$ measured at the LHC in
the forward rapidity region is a sensitive probe of the low-x dynamics and can help 
constrain $Q_{0s}^2$ further. We have shown that various theoretical uncertainties 
in our formalism are minimized at very forward rapidities at the LHC. Therefore, 
measuring the nuclear modification factor $R_{pA}$ in the {\it very forward region} in 
proton-nucleus collisions at the LHC will be a robust test of gluon saturation
dynamics and the Color Glass Condensate formalism.

\begin{acknowledgments}
We would like to thank David d'Enterria, Adrian Dumitru, Alex Kovner,
Anna Stasto, Mark Strikman and Dionisis Triantafyllopoulos for useful
discussions. We are grateful to the organizers of ``High-energy QCD
after the start of the LHC'' workshop in the Galileo Galilei Institute for Theoretical Physics (Florence) and
``Frontiers in QCD'' workshop in the  Institute for Nuclear Theory (Seattle) for their
hospitality and invitation to these stimulating workshops where this
paper was finalized.  J.J-M. is supported in part by the DOE Office of
Nuclear Physics through Grant No.\ DE-FG02-09ER41620, from the ``Lab
Directed Research and Development'' grant LDRD~10-043 (Brookhaven
National Laboratory), and from The City University of New York through
the PSC-CUNY Research Program, grant 64554-00 42. The work of A.H.R is
supported in part by Fondecyt grants 1110781.
\end{acknowledgments}


\end{document}